\documentclass[10pt,final,twocolumn]{IEEEtran}
\usepackage[utf8]{inputenc}
\usepackage{textcomp}
\DeclareSymbolFont{tildelow}{TS1}{cmr}{m}{n}
\DeclareMathSymbol{\tildelow}{0}{tildelow}{126}
\usepackage{bbm}
\usepackage{xcolor}
\usepackage{color}
\usepackage{graphicx}
\usepackage{epsfig}
\usepackage{subfigure}
\usepackage{amssymb}
\usepackage{amsbsy}
\usepackage{cite}
\usepackage{amsthm}
\usepackage{romannum}
\usepackage{algorithm}
\usepackage{algpseudocode}
\usepackage{verbatim}
\usepackage{stackengine}
\usepackage{amsmath}

\newtheorem{theorem}{Theorem}
\newtheorem{lemma}{Lemma}
\newtheorem{remark}{Remark}
\newtheorem{definition}{Definition}

\newcommand{\beq}{\begin{equation}}
\newcommand{\eeq}{\end{equation}}
\newcommand{\bea}{\begin{array}}
\newcommand{\ena}{\end{array}}
\newcommand{\bds}{\begin {itemize}}
\newcommand{\eds}{\end {itemize}}
\newcommand{\bdf}{\begin{definition}}
\newcommand{\blm}{\begin{lemma}}
\newcommand{\edf}{\end{definition}}
\newcommand{\elm}{\end{lemma}}
\newcommand{\bthm}{\begin{theorem}}
\newcommand{\ethm}{\end{theorem}}
\newcommand{\bprp}{\begin{prop}}
\newcommand{\eprp}{\end{prop}}
\newcommand{\bcl}{\begin{claim}}
\newcommand{\ecl}{\end{claim}}
\newcommand{\bcr}{\begin{coro}}
\newcommand{\ecr}{\end{coro}}
\newcommand{\bquest}{\begin{question}}
\newcommand{\equest}{\end{question}}

\newcommand{\larrow}{{\larrow}}

\def\urltilda{\kern -.15em\lower .7ex\hbox{\~{}}\kern .04em}

\usepackage{float}
\usepackage[T1]{fontenc}
\usepackage{url}
\usepackage{ifthen}
\usepackage{cite}

\renewcommand{\baselinestretch}{1.1}
\begin{document}

\title{Searching for a Hidden Markov Anomaly over
Multiple Processes}

\author{
  {Levli Citron, Kobi Cohen, Qing Zhao}
	\thanks{
	L. Citron and K. Cohen are with the School of Electrical and Computer Engineering, Ben-Gurion University of the Negev, Beer-Sheva, Israel (e-mail:levlic@post.bgu.ac.il; yakovsec@bgu.ac.il). 
	}
 \thanks{
	Q. Zhao is with the School of Electrical and Computer Engineering, Cornell University, Itaca, NY, US (e-mail: qz16@cornell.edu). 
	}
    	\thanks{The work of L. Citron and K. Cohen was supported by Binational Science Foundation (grant No. 2024611). The work of Q. Zhao was supported by the National Science Foundation under Grant CCF-2419622.}
    \thanks{A short version of this paper that introduces the ADHM algorithm, and preliminary simulation results was presented in Proceedings of the IEEE International Symposium on Information Theory (ISIT) 2024 \cite{citron2024anomaly}. This extended journal version significantly expands upon the earlier work and includes: (i) a detailed description of the ADHM algorithm and its extensions; (ii) a rigorous asymptotic analysis, including complete proofs that provide theoretical support for the design of ADHM; (iii) comprehensive simulation results across a broader range of scenarios; and (iv) an in-depth discussion of the results, including a thorough comparison with existing methods in the literature.}
    \thanks{This work has been submitted to the IEEE for possible publication. Copyright may be transferred without notice, after which this version may no longer be accessible.}
	\vspace{-0.75cm }
}
\maketitle
\pagenumbering{arabic}

\begin{abstract}

We address the problem of detecting an anomalous process among a large number of processes. At each time $t$, normal processes are in state zero (normal state), while the abnormal process may be in either state zero (normal state) or state one (abnormal state), with the states being hidden. The transition between states for the abnormal process is governed by a Markov chain over time. At each time step, observations can be drawn from a selected subset of processes. Each probed process generates an observation depending on its hidden state—either a typical distribution under state zero or an abnormal distribution under state one.

The objective is to design a sequential search strategy that minimizes the expected detection time, subject to an error probability constraint. In contrast to prior works that assume i.i.d. observations, we address a new setting where anomalies evolve according to a hidden Markov model. To this end, we propose a novel algorithm, dubbed Anomaly Detection under Hidden Markov model (ADHM), which dynamically adapts the probing strategy based on accumulated statistical evidence and predictive belief updates over hidden states. ADHM effectively leverages temporal correlations to focus sensing resources on the most informative processes. The algorithm is supported by an asymptotic theoretical foundation, grounded in an oracle analysis that characterizes the fundamental limits of detection under the assumption of a known distribution of the hidden states. In addition, the algorithm demonstrates strong empirical performance, consistently outperforming existing methods in extensive simulations.\vspace{0.2cm}

\end{abstract}

\vspace{0.0cm}
\section{Introduction}
\label{sec:introduction}

In many real-world systems, timely detection of rare or abnormal events is critical for ensuring efficiency, security, and resilience~\cite{cohen2015active, ho2015anomaly, hemo2020searching, song2017asymptotically, vaidhiyan2017learning, zhong2019deep,  tsopelakos2022sequential, kartik2022fixed, tsopelakos2023asymptotically}. Whether identifying intrusions in a computer network, detecting transmission opportunities in dynamic spectrum access, or monitoring critical infrastructure for signs of failure, decision-makers must act quickly based on limited and often noisy observations. This work addresses a fundamental instance of such detection problems.

We focus on the task of detecting an anomalous process among a collection of $M$ processes. Adopting terminology from target search, we refer to these processes as \textit{cells}, with the anomalous process representing the \textit{target}, which may be located in any of the $M$ cells. At each time step $t$, normal processes remain in a hidden state labeled as state zero (normal), while the anomalous process evolves according to a hidden Markov model (HMM) with two hidden states: state zero (normal) and state one (abnormal). The decision-maker is permitted to probe $K$ out of the $M$ cells at each time step ($1 \leq K \leq M$). The observation obtained from any probed process follows distribution $f$ when in state zero, and distribution $g$ when in state one.

This setting captures a prototypical scenario for detecting uncommon but significant events in large-scale systems. Such events may correspond to opportunities---such as spectrum availability in wireless communications---or threats, including malicious activities or system faults. Depending on the context, each cell may represent an autonomous process that emits an observation when probed, and the challenge lies in efficiently identifying the anomalous one with minimal delay and high reliability.

Unlike related studies on anomaly detection formulations, the temporal correlation introduced by the hidden Markov model adds a layer of complexity to the search process. This makes the setting more challenging and distinct from standard models that assume independent observations. In the following sections, we describe our approach to solving this problem and outline the main contributions of this work.

\subsection{Searching for Anomalies over Multiple Processes}

In sequential search problems involving a finite set of processes, a central challenge lies in determining which subset of processes to probe at each time step in order to efficiently and reliably detect the presence and location of an anomaly. This class of problems falls under the broader framework of controlled sensing for hypothesis testing~\cite{nitinawarat2013controlled, nitinawarat2015controlled}, also known as active hypothesis testing (AHT)~\cite{naghshvar2013active, naghshvar2013sequentiality}. These frameworks trace their roots to Chernoff’s seminal work on the sequential design of experiments~\cite{chernoff1959sequential}, which focused on binary hypothesis testing under i.i.d. observations and introduced the renowned Chernoff test—an asymptotically optimal randomized strategy as the error probability tends to zero.

Over the years, Chernoff’s foundational ideas have been significantly extended and generalized to handle a wide range of sequential decision-making scenarios. Notable extensions include multihypothesis testing, structured sampling constraints, and various cost models~\cite{mukherjee2022active, cohen2015active, lambez2021anomaly, huang2018active, gurevich2019sequential, tsopelakos2022sequential, tsopelakos2019sequential, tsopelakos2023asymptotically, gafni2023anomaly, cohen2015asymptotically, song2017asymptotically, hemo2020searching, vaidhiyan2017learning, kaspi2017searching, zhong2019deep, tajer2021active, kartik2022fixed, vershinin2025multi}. In particular, the problem of anomaly detection over multiple processes has attracted growing attention, as seen in~\cite{cohen2015active, lambez2021anomaly, huang2018active, gurevich2019sequential, tsopelakos2022sequential, tsopelakos2019sequential, tsopelakos2023asymptotically, gafni2023anomaly, cohen2015asymptotically, song2017asymptotically, hemo2020searching, vaidhiyan2017learning, zhong2019deep, kartik2022fixed}. In our prior work~\cite{cohen2015active}, we proposed a deterministic sampling policy for active hypothesis testing in the context of anomaly detection with i.i.d. observations. Subsequent extensions addressed practical challenges such as minimizing switching costs~\cite{lambez2021anomaly}, handling heterogeneous observation models~\cite{huang2018active}, and incorporating nonlinear cost structures~\cite{gurevich2019sequential}. Other studies have explored related directions including anomaly detection with constraints on the number of anomalies~\cite{song2017asymptotically}, settings involving Poisson processes~\cite{vaidhiyan2017learning}, dynamically varying probing capacity~\cite{tsopelakos2022sequential}, and hierarchical sampling strategies~\cite{wang2020information, gafni2021searching, gafni2023anomaly}.

Despite this extensive body of work, most existing approaches assume that observations are i.i.d. across time. This assumption, while simplifying analysis, limits applicability to settings where observations exhibit temporal dependence. A notable exception is~\cite{nitinawarat2015controlled}, which extends Chernoff's framework to handle observations governed by a stationary, discrete Markov model. However, that work does not consider hidden Markov dynamics or continuous observation spaces.

In contrast, the current paper addresses a more complex and realistic setting where the anomalous process evolves according to an HMM with two latent states representing normal and abnormal behavior. This introduces temporal correlation in the observations and poses new challenges for sequential decision-making. Moreover, the proposed framework accommodates both discrete and continuous observations, further broadening its applicability to real-world detection tasks.

\subsection{Main Results}

We introduce a novel anomaly detection model within the framework of active hypothesis testing. While recent works have developed closed-loop selection algorithms—where the choice of which cells to observe is guided by past observations—for known and general distributions \cite{cohen2015active, vaidhiyan2017learning, naghshvar2013active, vaidhiyan2017learning}, we consider a new challenging setting. Specifically, we address the problem of detecting an anomalous cell among many, where the observations are temporally correlated and governed by an HMM. This scenario reflects realistic systems in which processes may switch between normal and abnormal behavior, either intentionally (e.g., due to adversarial attacks) or unintentionally (e.g., due to malfunctioning components). The HMM-based anomaly model presents unique challenges that necessitate the development of new algorithmic techniques.

On the algorithmic front, we propose a novel approach tailored to the HMM-based anomaly detection setting, termed \textit{Anomaly Detection under Hidden Markov model (ADHM)}. In ADHM, the selection rule dynamically identifies the $K$ cells most likely to contain the anomaly to observe at time $n$, based on the evolving belief over the system's hidden states. The algorithm leverages the temporal structure of the HMM to track the likelihood of the anomaly’s location and its underlying distribution. At each time step, ADHM computes the forward variables for the HMM using past observations, the transition matrix, and initial state probabilities. These forward variables capture the probability of each hidden state at time $n$, enabling the algorithm to continuously update its estimate of the anomalous distribution. From a performance analysis perspective, the design of ADHM is based on asymptotic analysis in the vanishing error probability regime, achieving optimal performance under the oracle assumption of known hidden state distributions. Furthermore, simulation results demonstrate that ADHM yields substantial performance gains over existing methods, even in practical finite-sample settings.

\subsection{Other Related Work}

%SPRT
Sequential search for hypothesis testing have garnered significant attention ever since Wald's work on sequential analysis \cite{Wald:1947}. Wald presented the properties of sequential tests, allowing decisions to be reached much earlier than with fixed-size tests. He established the Sequential Probability Ratio Test (SPRT), which is optimal in temrs of minimizing the detection delay under false alarm and missed detection constraints. Various extensions for multi-hypothesis testing have been studied in \cite{mukherjee2023sprt,bussgang1970sequential,boggio2003active,dragalin1996simple,tartakovsky2002efficient,xing2023signal,tartakovsky2021optimal,lai2011quickest,malloy2014sequential,malloy2012quickest}. Chernoff's In this paper, a modified SPRT for the decision rule is established while only considering the probability of false alarm. Unlike in \cite{Wald:1947}, we consider a constrained observation model. Chernoff’s sequential design
of experiments extended this problem by incorporating a control aspect on the observation distribution \cite{chernoff1959sequential, nitinawarat2013controlled, naghshvar2013active}. 

The studies on sequential search are extensive, and the problem of detecting the first anomaly process among an infinite number of i.i.d. processes ($M \rightarrow \infty$) was explored in \cite{tajer2013quick,lai2011quickest,malloy2012quickest,malloy2014sequential, xing2023signal}. In this case, the solution involves executing an independent sequential test without memory for each individual process. When the decision maker switches to a different process, the selection of the new process is arbitrary, and a new sequential test starts afresh. Unlike in this paper, where a finite (potentially large) number of processes is known and there is a single target. Consequently, the states of the processes exhibit correlation in this model. In such a framework, the choice of the selection rule determining which process to observe at each time becomes crucial for minimizing detection delay. This contrasts with \cite{tajer2013quick, lai2011quickest, malloy2012quickest, malloy2014sequential,  xing2023signal}, where the order of process observation is deemed irrelevant. 

An alternative line of research leverages deep reinforcement learning (DRL), which has gained traction in recent years as a means of addressing active statistical inference problems~\cite{kartik2018policy, puzanov2018deep, zhong2019deep, puzanov2020deep, joseph2020anomaly, kalouptsidis2025neural}. Unlike the analytic, asymptotic methods traditionally used in AHT, DRL-based approaches offer approximate solutions in scenarios where tractable, structured policies are difficult to design. However, this flexibility comes at the cost of significantly higher complexity, both in terms of offline training and real-time execution. Despite these challenges, certain studies have demonstrated that efficient implementations are possible even on low-cost hardware, making DRL practical for real-world deployment~\cite{livne2020pops}. Specifically, DRL algorithms tailored for single-agent AHT problems have been proposed in~\cite{zhong2019deep, joseph2020anomaly, kartik2018policy, zhong2022controlled}, highlighting the potential of DRL to effectively handle decision-making under uncertainty in such settings. For instance, in~\cite{kartik2018policy}, a Q-learning approach was developed to construct an experiment selection strategy that maximizes the rate of increase in confidence toward the correct hypothesis. Actor-Critic methods were employed in~\cite{zhong2019deep, joseph2020anomaly} to achieve similar goals. More recently, multi-agent extensions of DRL for AHT problems have emerged, focusing on scalable and decentralized learning architectures. Contributions in this direction include~\cite{szostak2022decentralized, stamatelis2023deep, joseph2023scalable, szostak2024deep}, where agents collaboratively learn efficient policies in distributed environments.

\subsection{Organization}

The remainder of the paper is organized as follows. Section~\ref{sec:System_Model} introduces the system model and problem formulation. In Section~\ref{sec:ADHM}, we present the ADHM algorithm, its theoretical foundation, and proposed extensions. Section~\ref{sec:simulations} provides numerical evaluations of the ADHM algorithm, and Section~\ref{sec:Conclusion} concludes the paper.

\section{System Model and Problem Formulation}
\label{sec:System_Model}

We address the problem of sequentially detecting an anomalous process (referred to as the \emph{target}) among a large set of $M$ processes, each associated with a distinct cell. If the target resides in cell $m$, the true hypothesis is denoted by $H_m$. The prior probability that hypothesis $H_m$ is true is given by $\pi_m$, where the probabilities satisfy $\sum_{m=1}^M \pi_m = 1$ and $0 < \pi_m < 1$ for all $m$, to avoid trivial solutions.

\begin{figure}[htbp]
    \centering
    \includegraphics[scale=0.5]{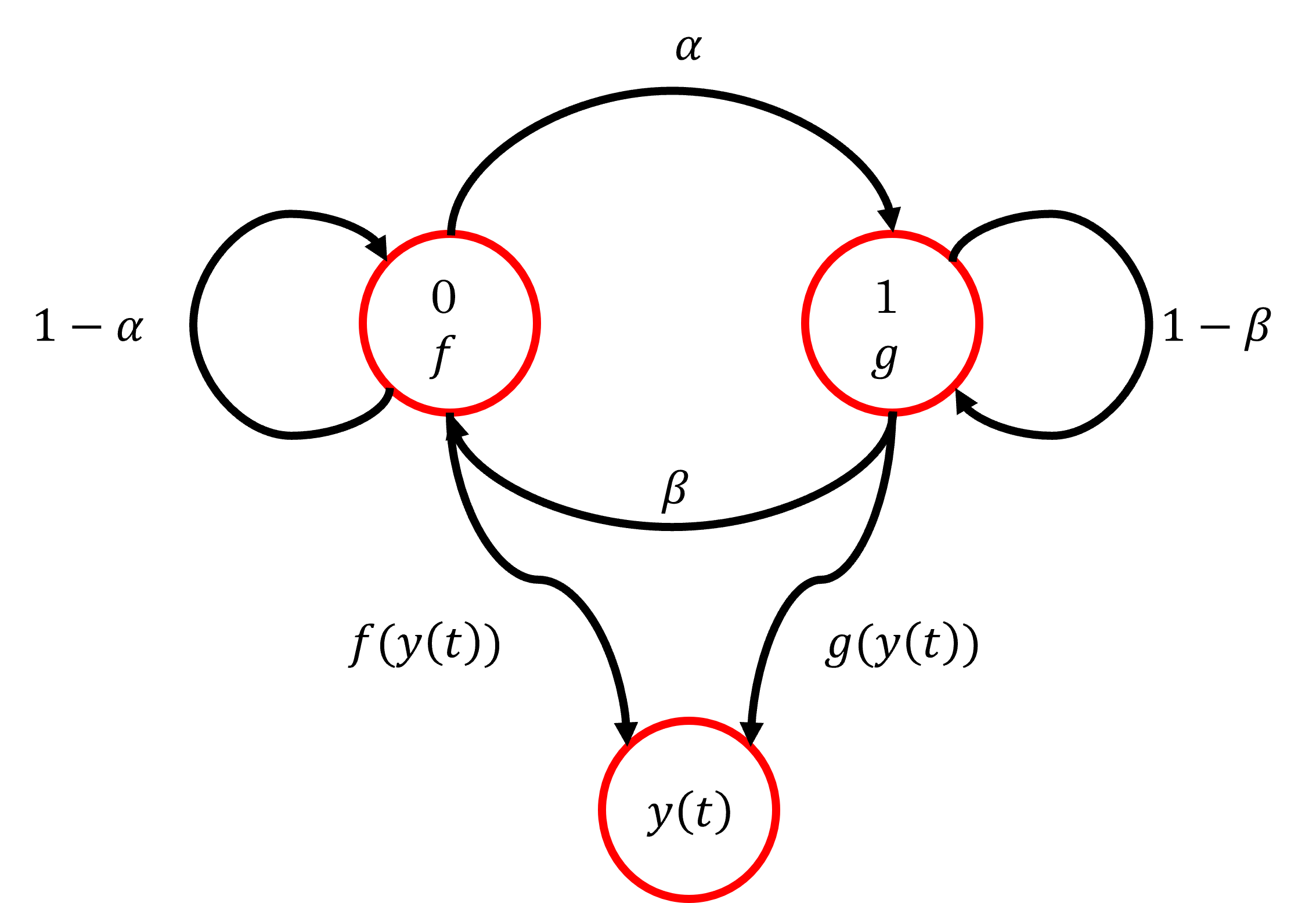}
    \caption{An illustration of the observation process of the anomalous cell as governed by the HMM. When the anomalous process is selected for observation at time $t$, a measurement $y(t)$ is obtained, which follows distribution $f$ or $g$ depending on the current hidden state of the Markov chain—state~$0$ (normal) or state~$1$ (abnormal), respectively.}
    \label{fig:HMM}
\end{figure}

At each time step, observations can be observed from a subset of $K$ cells, with $1 \leq K \leq M$. When cell $m$ is observed at time $n$, an observation $y_m(n)$ is obtained. If the hypothesis $H_m$ is incorrect (i.e., the target is not in cell $m$), the observation $y_m(n)$ follows a distribution $f(y)$. If $H_m$ is correct, the sequence of observations $\{y_m(n)\}$ is generated by an HMM. The HMM consists of two hidden states: state~0 (normal), associated with the distribution $f(y)$, and state~1 (abnormal), associated with the distribution $g(y)$. At each time, the observation $y_m(n)$ is drawn according to the current hidden state of the HMM. The state transitions are governed by a known Markov transition matrix, denoted by
\begin{equation}
\label{eq:transition matrix}\textbf{A}=\left( \begin{array}{cc} 1-\alpha & \alpha \\
\beta & 1-\beta \end{array}\right).\end{equation}
Here, $\alpha$ represents the probability of transitioning from state 0 to state 1, while $\beta$ represents the probability of transitioning from state 1 to state 0.
The stationary probability of being in state zero is given by: $\pi^0=\frac{\beta}{\alpha+\beta}$. Let $P^0_t$ be the probability of the HMM being in state zero at time $t$ based on the history of the process. Then, if hypothesis $H_m$ is true, $y_m(t)$ follows distribution 
\begin{equation}
 \sim\tilde{g}_r\triangleq r f(y)+(1-r) g(y),
\end{equation}
conditioned on $P^0_t=r$, $0\leq r\leq 1$.
Figure \ref{fig:HMM} illustrates the observation process of the anomalous cell as governed by the HMM model.

We aim at developing an active search strategy, denoted by $\Gamma = (\phi, \tau, \delta)$, where $\tau$ is a stopping time that determines when to terminate the search. At each time step $1 \leq n < \tau$, the action rule $\phi$ specifies which subset of cells to observe. Upon stopping at time $\tau$, a terminal decision rule $\delta \in \{1, 2, \ldots, M\}$ is applied to declare the estimated location of the target.

Let $\mathbf{P}_m$ denote the probability measure under hypothesis $H_m$, and let $\mathbf{E}_m$ represent the corresponding expectation operator. The error probability under strategy $\Gamma$ is given by:
\begin{equation}
\begin{aligned}
\label{Prob.error}
P_e(\Gamma) &\triangleq \sum_{m=1}^M \pi_m \alpha_m, \\
\end{aligned}
\end{equation}
where $\alpha_m = \textbf{P}_m(\delta \neq m | \Gamma)$ is the probability of declaring $\delta \neq m$ when $H_m$ is true.

The sample complexity, representing the average delay under strategy $\Gamma$, is given by:
\begin{equation}
\begin{aligned}
\label{SampleComp}
\textbf{E}[\tau|\Gamma] &\triangleq \sum_{m=1}^M \pi_m \textbf{E}_m[\tau|\Gamma].
\end{aligned}
\end{equation}

We adopt a Bayesian approach, as is widely used in active hypothesis testing (e.g., \cite{chernoff1959sequential, cohen2015active}), and define the Bayes risk accordingly. Specifically, each observation incurs a sampling cost of $c$, and a unit loss is incurred for an incorrect final decision. Thus, the Bayes risk associated with a strategy $\Gamma$ is given by:
\begin{equation}
\begin{aligned}
\label{BayesRiskm}
R_m(\Gamma) &\triangleq \alpha_m(\Gamma) + c \cdot \textbf{E}_m(\tau|\Gamma).
\end{aligned}
\end{equation}

The average Bayes risk is given by:
\begin{equation}
\begin{aligned}
\label{BayesRisk}
R(\Gamma) &\triangleq P_e(\Gamma) + c \cdot \textbf{E}(\tau|\Gamma).
\end{aligned}
\end{equation}

The objective is to derive the strategy $\Gamma$ that minimizes the Bayes risk $R(\Gamma)$. This leads to the following optimization problem:
\begin{equation}
\begin{aligned}
\label{Optimize}
\inf_{\Gamma} R(\Gamma).
\end{aligned}
\end{equation}

\section{The Anomaly Detection under Hidden Markov model (ADHM) Algorithm}
\label{sec:ADHM}

We now present the \textit{Anomaly Detection under Hidden Markov model (ADHM)} algorithm to address \eqref{Optimize}. 

We begin by introducing the notations used in ADHM. Let
\begin{equation}
\label{eq:LLR}
\ell_m(n) \triangleq \log \frac{\tilde{g}_r(y_m(n))}{f(y_m(n))},
\end{equation}
denote the log-likelihood ratio (LLR) for cell \( m \) at time \( n \), conditioned on the belief \( P^0_n = r \). The accumulated observed LLRs for each cell are defined by
\begin{equation}
\label{eq:SUM_LLR}
S_m(n) \triangleq \sum_{t=1}^{n} \ell_m(t) \cdot \mathbf{1}_m(t),
\end{equation}
where \( \mathbf{1}_m(t) = 1 \) if cell \( m \) was observed at time \( t \), and 0 otherwise.

Let \( m^{(i)}(n) \) denote the index of the cell with the \( i \)-th largest observed sum LLR at time \( n \). The difference between the top two observed sum LLRs is defined as
\begin{equation}
\label{eq:SLLR_dif}
\Delta S(n) \triangleq S_{m^{(1)}(n)}(n) - S_{m^{(2)}(n)}(n).
\end{equation}

\subsection{The ADHM Algorithm}

The ADHM algorithm initializes at time \( n = 1 \) with \( S_m(1) = 0 \) for all \( m \in \{1, \dots, M\} \), and with prior belief \( P^0_1 = \beta / (\alpha + \beta) \). At each time step \( n \), the algorithm selects the \( K \) cells with the highest sum LLRs:
\begin{equation}
\label{eq:selection_rule}
\phi(n) = \left( m^{(1)}(n), m^{(2)}(n), \ldots, m^{(K)}(n) \right),
\end{equation}
where ties can be resolved arbitrarily. The selected cells are observed, their LLRs are computed using the current belief \( P^0_n \), and their cumulative LLRs are updated accordingly.

Following the observation, the belief \( P^0_{n+1} \), representing the probability that the system is in the normal state at time \( n+1 \), is updated using the forward recursion of the HMM. Specifically, let \( y(n) \) denote the set of observed values at time \( n \). Then, define:\begin{equation}
\label{eq:belief_N_D}
\begin{aligned}
\mathcal{N}(n) &\triangleq P^0_n \cdot (1 - \alpha) \cdot f(y(n)) + (1 - P^0_n) \cdot \beta \cdot g(y(n)), \\
\mathcal{D}(n) &\triangleq P^0_n \cdot  f(y(n)) + (1 - P^0_n) \cdot g(y(n)), 
\end{aligned}
\end{equation}
and compute the updated belief as:
\begin{equation}
\label{eq:HMM_update}
P^0_{n+1} = \frac{\mathcal{N}(n)}{\mathcal{D}(n)}.
\end{equation}
The algorithm continues until the stopping condition:
\begin{equation}
\label{eq: stopping rule}
\tau = \inf \left\{ n \;:\; \Delta S(n) \geq -\log c \right\}
\end{equation}
is satisfied, at which point the final decision is to select the cell with the largest accumulated LLR:
\begin{equation}
\label{eq: decision rule}
\delta = m^{(1)}(\tau).
\end{equation}

The design of ADHM is grounded in asymptotic analysis, where the goal is to solve \eqref{Optimize} as the error probability tends to zero. While proving asymptotic optimality in full generality remains a challenge, we show that ADHM achieves asymptotically optimal performance under the oracle assumption of known hidden state distributions. That is, if the belief computed via the forward algorithm accurately tracks the true hidden state distribution, then the performance of ADHM matches that of an oracle that knows the true distributions. We refer to this idealized scenario as the \textit{oracle case}.

\begin{theorem}
\label{thm:adhm_optimality}
Under the oracle case, ADHM solves \eqref{Optimize} asymptotically as \( c \to 0 \). 
\end{theorem}

The proof can be found in the Appendix.

The ADHM algorithm can be intuitively explained as follows. At each time, the anomaly distribution changes according to the hidden states of the HMM. Notably, the algorithm aims at predicting the correct distribution at each time \( n \), denoted by \( \tilde{g}_n \). This is done using the forward recursion of the HMM. Then, the observations can be regarded as being drawn from the updated distribution, which is used in the computation of the sum LLRs for the \( K \) selected observed cells. Opting for the \( K \) cells with the highest sum LLRs ensures the extraction of the most informative data from those cells. A detailed pseudocode of ADHM is presented in Algorithm \ref{pseudo_code_1}.

\begin{algorithm}
\caption{ADHM Algorithm}
\label{pseudo_code_1}
\begin{algorithmic}[1]
\State \textbf{Input:} Number of observed cells \( K \), transition matrix \( \mathbf{A} \), confidence parameter \( c \).
\State \textbf{Output:} Selected target cell \( \delta \), stopping time \( \tau \).
\State Initialize \( S_m \gets 0 \) for all \( m = 1, \ldots, M \), and \( P^0_1 \gets \beta / (\alpha + \beta) \), set \( n \gets 1 \).
\While{true}
    \State Compute selection rule: \\
    \phantom{space}\hspace{2cm} \( \phi(n) \gets \{ m^{(1)}(n), \ldots, m^{(K)}(n) \} \)
    \State Observe outputs \( y_m(n) \) for all \( m \in \phi(n) \)
    \State Compute LLRs \( \ell_m(n) \) using \( P^0_n \)
    \State Update \( S_m(n) \gets S_m(n-1) + \ell_m(n) \) for observed \( m \)
    \State Compute \( \Delta S(n) \gets S_{m^{(1)}(n)}(n) - S_{m^{(2)}(n)}(n) \)
    \If{ \( \Delta S(n) \geq -\log c \) }
        \State \( \tau \gets n \), \( \delta \gets m^{(1)}(n) \)
        \State \textbf{break}
    \EndIf
    \State Update \( P^0_{n+1} \gets \mathcal{N}(n) / \mathcal{D}(n) \) using \eqref{eq:belief_N_D}
    \State \( n \gets n + 1 \)
\EndWhile
\State \Return \( \delta, \tau \)
\end{algorithmic}
\end{algorithm}

\subsection{ADHM with Predictive Time Scheduling (ADHM-P)}

In the original ADHM algorithm, sampling occurs at every time step, regardless of whether the resulting observation is likely to be informative. While this ensures continuous data collection, it can be inefficient—particularly when the hidden state of the anomaly is unlikely to change significantly over short time scales.

To improve sampling efficiency, we extend ADHM by incorporating \emph{predictive time scheduling}, resulting in the \emph{ADHM-P} algorithm. In this extended approach, sampling can be selectively skipped at time steps when the expected informativeness of the observation is low. The underlying HMM structure is exploited to predict the future evolution of the hidden states, enabling the algorithm to decide not only \emph{which} cells to observe, but also \emph{when} to observe them.

However, allowing the algorithm to skip sampling introduces an unintended consequence: In some cases, the policy may choose to observe too infrequently, particularly when the belief about the anomaly's location is not expected to change significantly. In such scenarios, the algorithm may defer sampling in anticipation of gathering more informative evidence in the future, potentially leading to increased detection delay. To prevent this, we introduce a \emph{cost per unit time} in addition to the existing cost per observation. This temporal cost discourages excessive delays in decision-making, ensuring a balance between timely detection and efficient sampling. 

In ADHM-P, the decision process is augmented with a \emph{scheduling rule} that determines whether to collect observations at each time step. Sampling is performed if the predicted informativeness—quantified by the posterior probability of being in the abnormal state—is sufficiently high. Additionally, to prevent long inactive periods that increase detection delay, we impose a safeguard condition that forces sampling if the delay exceeds a threshold. The pseudocode for the extended algorithm is presented in Algorithm \ref{pseudo_code_adhm_p}.

\begin{algorithm}
\caption{ADHM-P Algorithm}
\label{pseudo_code_adhm_p}
\begin{algorithmic}[1]
\State \textbf{Input:} Number of observed cells \( K \), transition matrix \( \mathbf{A} \), confidence parameter \( c \), scheduling threshold \( P_{\text{Th}} \), time-delay coefficient \( \gamma \).
\State \textbf{Output:} Selected target cell \( \delta \), stopping time \( \tau \).
\State Initialize \( S_m \gets 0 \) for all \( m = 1, \ldots, M \), \( P^0_1 \gets \beta / (\alpha + \beta) \), set \( n \gets 1 \), \( \tilde{N} \gets 0 \).
\While{true}
    \State Predict informativeness using belief: \( 1 - P^0_n \)
    \If{$(1 - P^0_n) > P_{\text{Th}}$ \textbf{or} \( \tilde{N} \cdot \gamma > c \)}
        \State Compute selection rule: \\
        \phantom{space}\hspace{2cm}\( \phi(n) \gets \{ m^{(1)}(n), \ldots, m^{(K)}(n) \} \)
        \State Observe outputs \( y_m(n) \) for all \( m \in \phi(n) \)
        \State Compute LLRs \( \ell_m(n) \) using \( P^0_n \)
        \State Update \( S_m(n) \gets S_{m}(n-1) + \ell_m(n) \) for \\
        \phantom{space}\hspace{0.3cm}observed \(  m \)
        \State Reset delay counter: \( \tilde{N} \gets 0 \)
    \Else
        \State Skip sampling; maintain current \\
        \phantom{space}\hspace{2cm}\( S_m(n) = S_m(n-1) \)
        \State Increment delay counter: \( \tilde{N} \gets \tilde{N} + 1 \)
    \EndIf
    \State Compute \( \Delta S(n) \gets S_{m^{(1)}(n)}(n) - S_{m^{(2)}(n)}(n) \)
    \If{ \( \Delta S(n) \geq -\log c \) }
        \State \( \tau \gets n \), \( \delta \gets m^{(1)}(n) \)
        \State \textbf{break}
    \EndIf
\If{sampling occurred at time \( n \)}
        \State Update \( P^0_{n+1} \gets \mathcal{N}(n) / \mathcal{D}(n) \) using observation\\ \phantom{space}\hspace{0.3cm} and transition model
    \Else
        \State Update \( P^0_{n+1} \) using only the transition matrix
    \EndIf
    \State \( n \gets n + 1 \)
\EndWhile
\State \Return \( \delta, \tau \)
\end{algorithmic}
\end{algorithm}

\subsection{Comparison with Existing AHT Methods}

A number of existing AHT algorithms, including the Chernoff test\cite{chernoff1959sequential}, and DGF policy\cite{huang2018active}, operate under the assumption that measurements are i.i.d. over time. While these methods often select actions based on sum LLRs, they do not exploit the temporal structure of the underlying process, particularly when the observation distribution evolves according to an HMM, as considered in this paper. In contrast, the ADHM algorithm is specifically designed to account for such temporal dependencies. Intuitively, ADHM, adapts to the dynamics of the system—leveraging structure to allocate probing effort more efficiently. This added structure allows ADHM to reduce uncertainty more aggressively, especially in settings where state evolution plays a key role. It does so by incorporating predictive models of the state evolution, allowing it to anticipate the distribution of future observations. This predictive capability enables more informed and forward-looking probe selection, leading to a faster accumulation of evidence in favor of the true hypothesis. 

Analytically, we show in the proof of Theorem~1 that, under the oracle case, ADHM achieves an asymptotic Bayes risk of $-c \log c / I^*_{\mathrm{ADHM}}$, where $I^*_{\mathrm{ADHM}}$ is the corresponding rate function, defined in~\eqref{eq:rate_function}. The following theorem establishes that this rate function is greater than or equal to that achieved by the Chernoff test (or equivalently, by DGF, which achieves the same rate).

\begin{theorem}
\label{thm:adhm_vs_dgf}
Let the achievable rate functions under ADHM and Chernoff algorithms be denoted by $I^*_{\mathrm{ADHM}}$ and $I^*_{\mathrm{Chernoff}}$, respectively. Then, under the oracle case, it holds that $I^*_{\mathrm{ADHM}} \geq I^*_{\mathrm{Chernoff}}$.
\end{theorem}

The proof can be found in the Appendix.

\noindent This result highlights the performance gain of ADHM in structured environments. By exploiting the temporal correlation and modeling the evolution of the underlying process, ADHM achieves a larger exponential decay rate in the probability of error, compared to policies like the Chernoff test (or DGF) that assume i.i.d. measurements. While the oracle case is considered for analytical tractability, simulation results show that ADHM consistently outperforms existing methods in practical scenarios where the true distributions over the hidden states are unknown.

\section{Simulation Results}
\label{sec:simulations}

In this section, we present numerical results that illustrate the performance of the proposed ADHM algorithm in the finite-sample regime. We compare its performance against two established benchmarks: the randomized Chernoff test \cite{chernoff1959sequential} and the deterministic DGF algorithm \cite{cohen2015active}.

We consider a simulation scenario involving a single anomalous object (referred to as the target) randomly located in one of \( M \) cells. The prior distribution is uniform, i.e., the probability that the target is in cell \( m \) is \( \pi_m = 1/M \) for all \( m = 1, \ldots, M \). At each time step \( n \), an observation \( y_m(n) \) is obtained from each selected cell. If cell \( m \) is normal (i.e., does not contain the target), the observation follows the nominal distribution \( f \). If the target is present in cell \( m \), then the observation \( y_m(n) \) is governed by a two-state HMM: state 0 corresponds to the normal distribution \( f \), while state 1 corresponds to the anomalous distribution \( g \). The underlying state transitions are defined by a known transition probability matrix $A$, as specified in~\eqref{eq:transition matrix}.
Let \( R_{\text{ADHM}} \), \( R_{\text{DGF}} \), and \( R_{\text{Ch}} \) denote the Bayes risk incurred under the ADHM algorithm, the DGF policy, and the Chernoff test, respectively. 

\begin{figure}[htbp]
    \centering
    \includegraphics[scale=0.3]{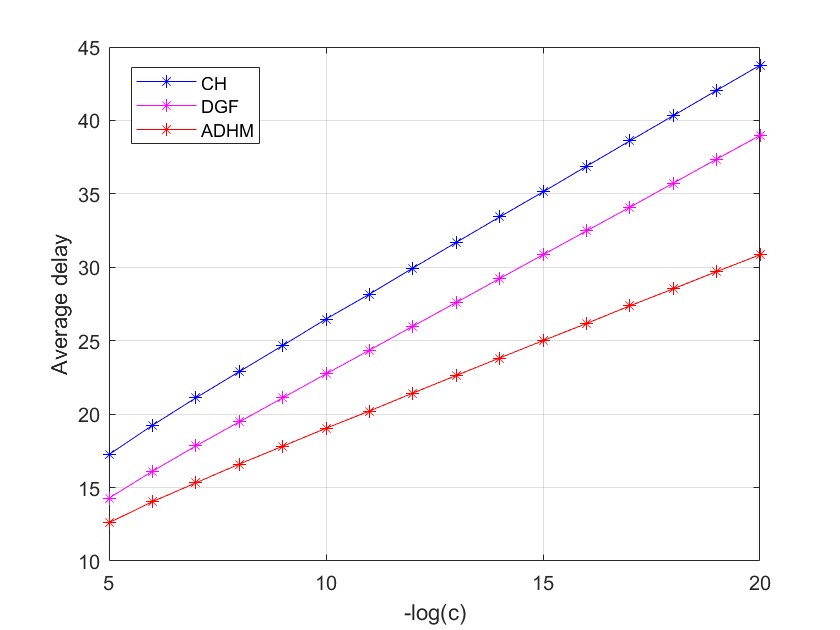}
    \caption{Performance comparison of three algorithms: the Chernoff test (CH), the DGF policy (DGF), and the proposed ADHM algorithm (ADHM), under the parameters \( M = 10 \), \( K = 2 \), \( \lambda_f = 0.5 \), \( \lambda_g = 10 \), and \( \alpha = \beta = 0.9 \). Results are based on \( 10^6 \) Monte Carlo trials. Shown: average detection delay versus \( -\log c \).}
    \label{fig:AD}
\end{figure}

\begin{figure}[htbp]
    \centering
    \includegraphics[scale=0.3]{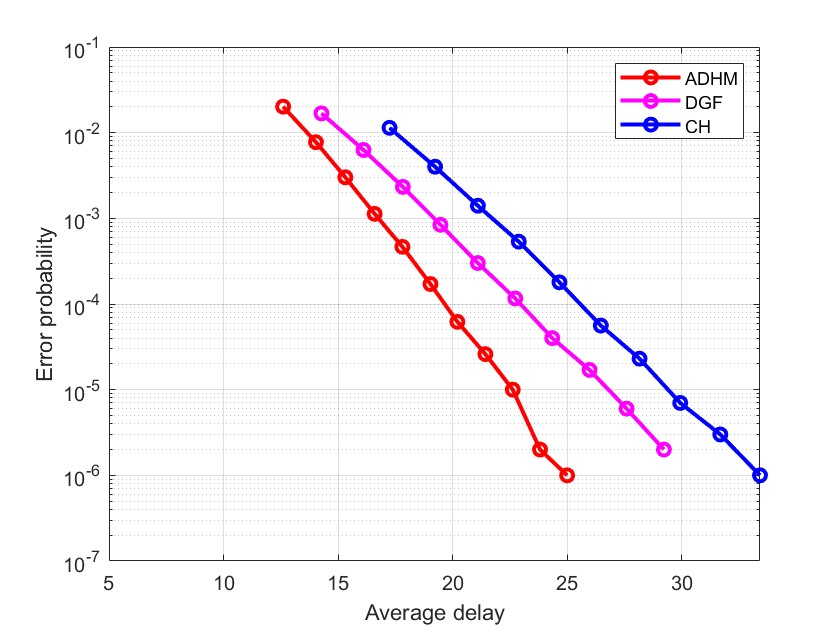}
    \caption{Performance comparison of three algorithms: the Chernoff test (CH), the DGF policy (DGF), and the proposed ADHM algorithm (ADHM), under the parameters \( M = 10 \), \( K = 2 \), \( \lambda_f = 0.5 \), \( \lambda_g = 10 \), and \( \alpha = \beta = 0.9 \). Results are based on \( 10^6 \) Monte Carlo trials. Shown: error probability versus average detection.}
    \label{fig:AD_vs_PE}
\end{figure}

\begin{figure}[htbp]
    \centering
    \includegraphics[scale=0.3]{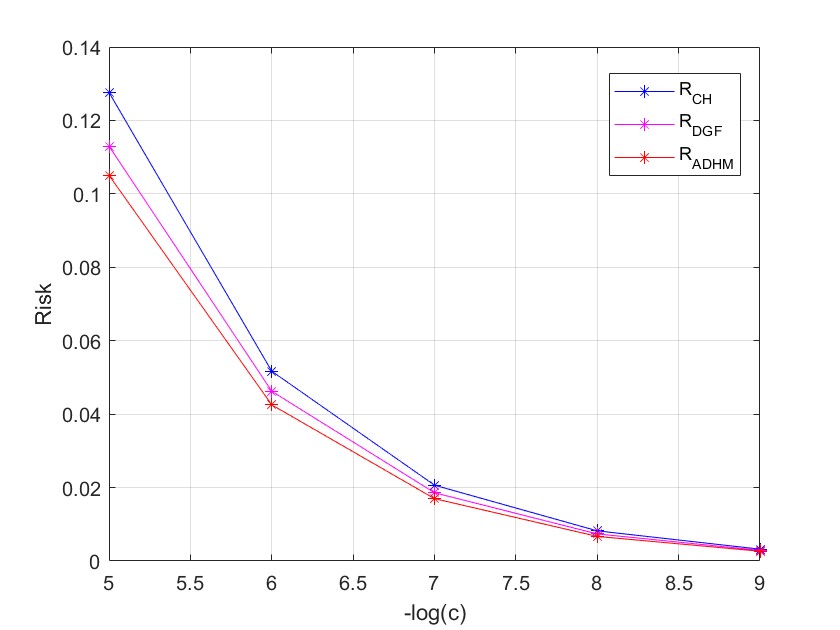}
    \caption{Performance comparison of three algorithms: the Chernoff test (CH), the DGF policy (DGF), and the proposed ADHM algorithm (ADHM), under the parameters \( M = 10 \), \( K = 2 \), \( \lambda_f = 0.5 \), \( \lambda_g = 10 \), and \( \alpha = \beta = 0.9 \). Results are based on \( 10^6 \) Monte Carlo trials. Shown: Bayes risk versus \( -\log c \).}
    \label{fig:Risk}
\end{figure}

We begin by evaluating the performance of the proposed algorithm in a setting with \( M = 10 \) cells, where at each time step \( K = 2 \) cells are observed. The parameters of the underlying HMM are set to \( \alpha = 0.9 \) and \( \beta = 0.9 \), indicating moderately persistent state transitions. The observation distributions are exponential, with \( f \sim \exp(\lambda_f) \) and \( g \sim \exp(\lambda_g) \), where the rate parameters are \( \lambda_f = 0.2 \) for the normal state and \( \lambda_g = 10 \) for the abnormal state, yielding a significant contrast between the two. The performance of the ADHM algorithm under this configuration is illustrated in Fig.~\ref{fig:AD}, Fig.~\ref{fig:AD_vs_PE}, and Fig.~\ref{fig:Risk}. Each result is based on \( 10^6 \) Monte Carlo trials to ensure statistical reliability.

Across all three performance metrics—average detection delay versus \(-\log c\) (Fig.~\ref{fig:AD}), error probability versus average delay (Fig.~\ref{fig:AD_vs_PE}), and Bayes risk versus \(-\log c\) (Fig.~\ref{fig:Risk})—the ADHM algorithm consistently outperforms both the deterministic DGF policy and the randomized Chernoff test. In particular, ADHM achieves the lowest detection delay for any given cost parameter or error tolerance, demonstrating its ability to exploit the hidden Markov dynamics effectively. The DGF policy yields intermediate performance, reflecting its use of past observations in deterministic selection decisions without fully leveraging temporal correlations of the HMM, while the Chernoff test exhibits the highest delay and risk, confirming that randomized sampling strategies are suboptimal in the HMM setting. These results underscore the benefits of adaptive, belief‐driven sampling in anomaly detection under HMM.

Next, we consider a second experimental setup with \( M = 10 \) cells and \( K = 2 \) observations per time step. The underlying HMM parameters are set to \( \alpha = 0.9 \) and \( \beta = 0.9 \), indicating strong persistence in both states. In this scenario, the observation model is discrete: observations follow a geometric distribution, with \( f \sim \text{Geometric}(\theta_f) \) and \( g \sim \text{Geometric}(\theta_g) \), where the success probabilities are set to \( \theta_f = 0.8 \) for the normal state and \( \theta_g = 0.5 \) for the abnormal state. This choice of parameters ensures a noticeable shift between the two distributions, even within the discrete domain.

\begin{figure}[htbp]
    \centering
    \includegraphics[scale=0.28]{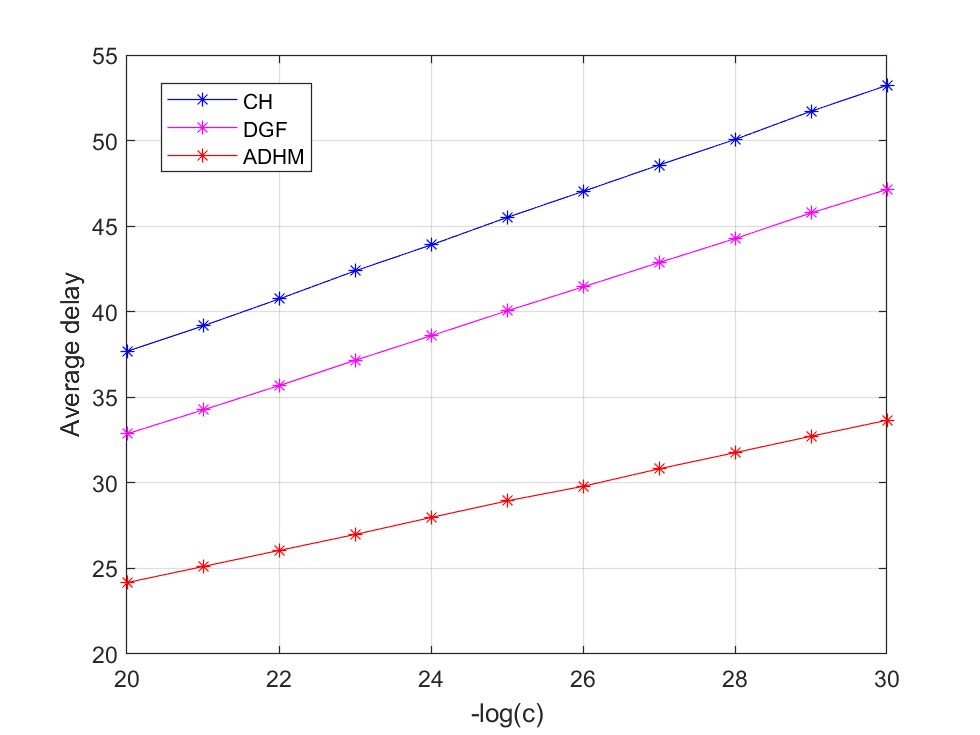}
    \caption{Performance comparison of three algorithms: the Chernoff test (CH), the DGF policy (DGF), and the proposed ADHM algorithm (ADHM), under the parameters \( M = 10 \), \( K = 5 \), \( \theta_f = 0.1 \), \( \theta_g = 0.9 \), and \( \alpha = \beta = 0.9 \). Results are based on \( 10^6 \) Monte Carlo trials. Shown: average detection delay versus \( -\log c \).}
    \label{fig:AD11}
\end{figure}

\begin{figure}[htbp]
    \centering
    \includegraphics[scale=0.28]{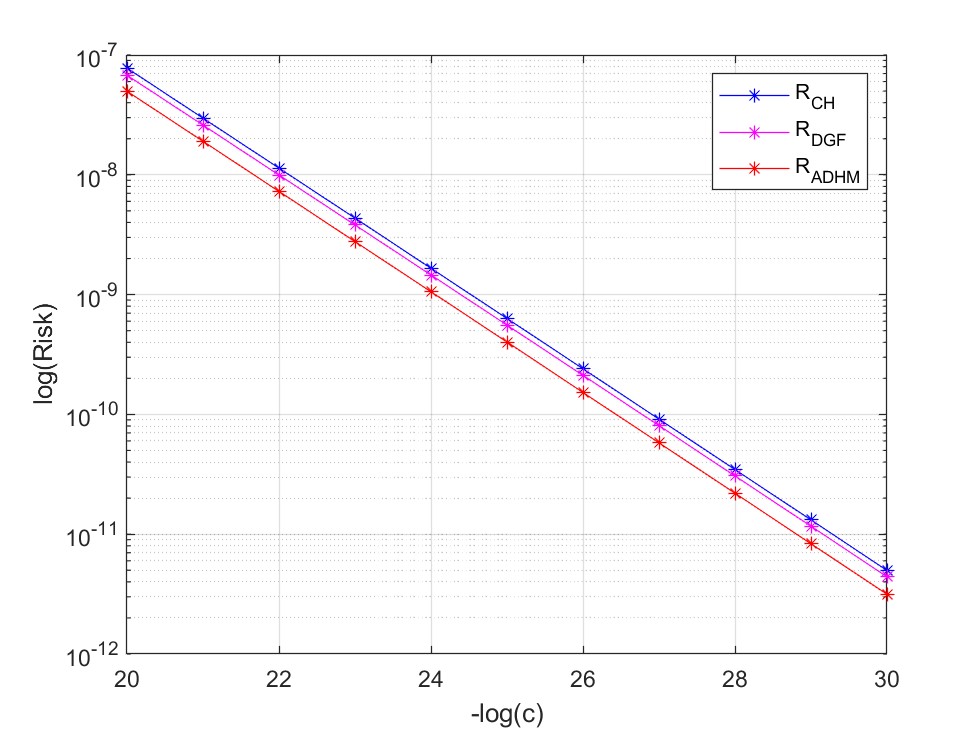}
    \caption{Performance comparison of three algorithms: the Chernoff test (CH), the DGF policy (DGF), and the proposed ADHM algorithm (ADHM), under the parameters \( M = 10 \), \( K = 5 \), \( \theta_f = 0.1 \), \( \theta_g = 0.9 \), and \( \alpha = \beta = 0.9 \). Results are based on \( 10^6 \) Monte Carlo trials. Shown: Bayes risk versus \( -\log c \).}
    \label{fig:Risk11}
\end{figure}

The performance of the ADHM algorithm in this setting is shown in Fig.~\ref{fig:AD11} and Fig.~\ref{fig:Risk11}, which display the average detection delay and the Bayes risk, respectively. Each curve is computed from \( 10^6 \) Monte Carlo trials to ensure robust statistical significance.

Despite the change in the observation model from continuous to discrete, the ADHM algorithm continues to exhibit clear advantages. It significantly outperforms both the deterministic DGF policy and the Chernoff test, further demonstrating its robustness and adaptability to different distribution families. These results emphasize the flexibility of ADHM and its effectiveness in leveraging temporal dynamics across a variety of observation models.

Finally, we compare the performance of the baseline ADHM algorithm with its extended version, ADHM-P, which incorporates predictive time scheduling. This evaluation highlights the impact of intelligently skipping uninformative observations on overall detection performance.

The comparison is conducted under two observation regimes: in the first, observations are collected at every time step, following the standard ADHM procedure; in the second, observations are selectively acquired under ADHM-P, which samples only when the predicted belief that the anomaly resides in the abnormal state exceeds a predefined threshold. This threshold is denoted by \( P_{\text{Th}} = 0.7 \), and serves as a condition to initiate sampling based on the current posterior belief. In addition, ADHM-P includes a mechanism to prevent excessive delay by forcing sampling when the system has remained idle for too long, as described earlier.

The simulation setup includes \( M = 5 \) cells and a budget of \( K = 2 \) observations per time step. The hidden Markov model parameters are set to \( \alpha = 0.1 \) and \( \beta = 0.1 \), representing rapidly switching behavior. Observations follow exponential distributions, with \( f \sim \exp(\lambda_f) \) and \( g \sim \exp(\lambda_g) \), where \( \lambda_f = 0.5 \) and \( \lambda_g = 10 \), ensuring a strong contrast between the normal and abnormal states.

\begin{figure}[htbp]
    \centering
    \includegraphics[scale=0.3]{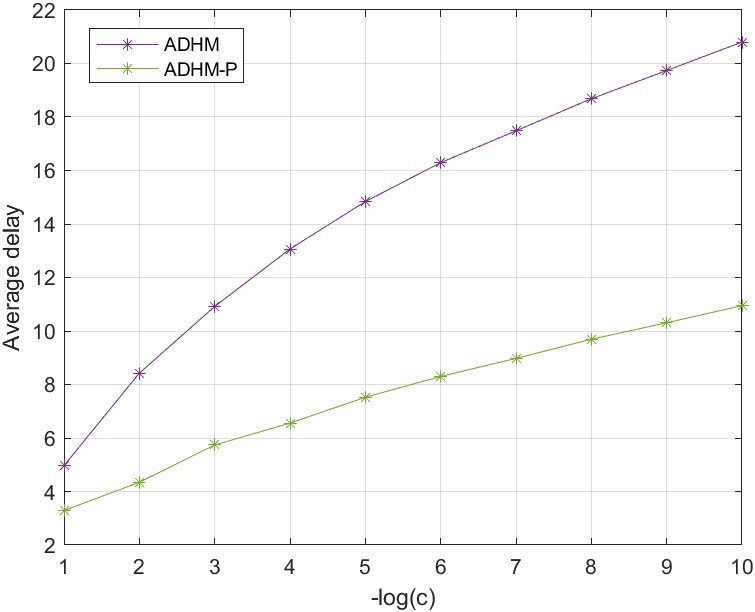}
    \caption{Performance comparison of two proposed algorithms: ADHM algorithm (ADHM), and ADHM with Predictive time scheduling (ADHM-P) for \( M = 5 \), \( K = 2 \), \( \lambda_f = 0.5 \), \( \lambda_g = 10 \), \( \alpha = \beta = 0.1 \), and threshold \( P_{\text{Th}} = 0.7 \). Results are based on \( 10^6 \) Monte Carlo trials. Shown: average detection delay versus \( -\log c \).}
    \label{fig:AD3}
\end{figure}

\begin{figure}[htbp]
    \centering
    \includegraphics[scale=0.3]{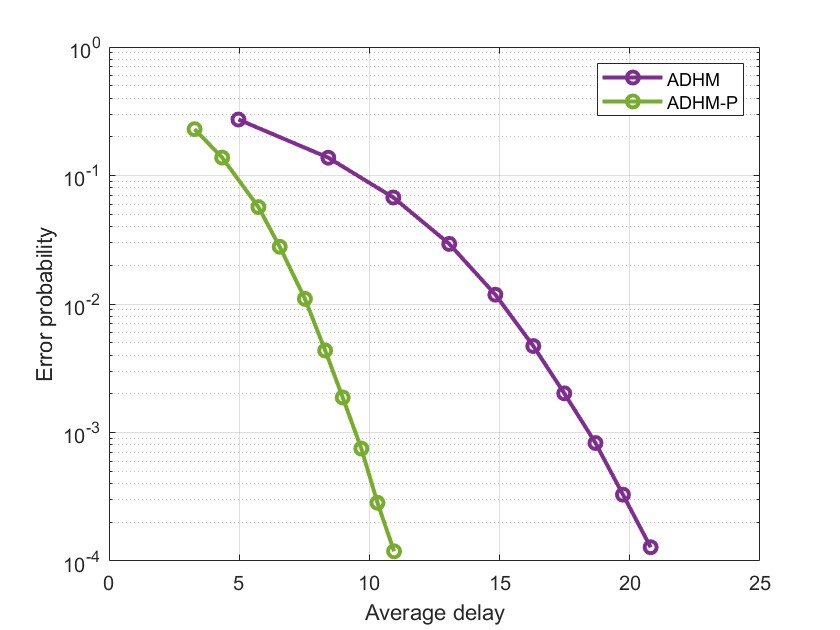}
    \caption{Performance comparison of two proposed algorithms: ADHM algorithm (ADHM), and ADHM with Predictive time scheduling (ADHM-P) for \( M = 5 \), \( K = 2 \), \( \lambda_f = 0.5 \), \( \lambda_g = 10 \), \( \alpha = \beta = 0.1 \), and threshold \( P_{\text{Th}} = 0.7 \). Results are based on \( 10^6 \) Monte Carlo trials. Shown: error probability versus average detection delay.}
    \label{fig:AD_vs_PE3}
\end{figure}

\begin{figure}[htbp]
    \centering
    \includegraphics[scale=0.3]{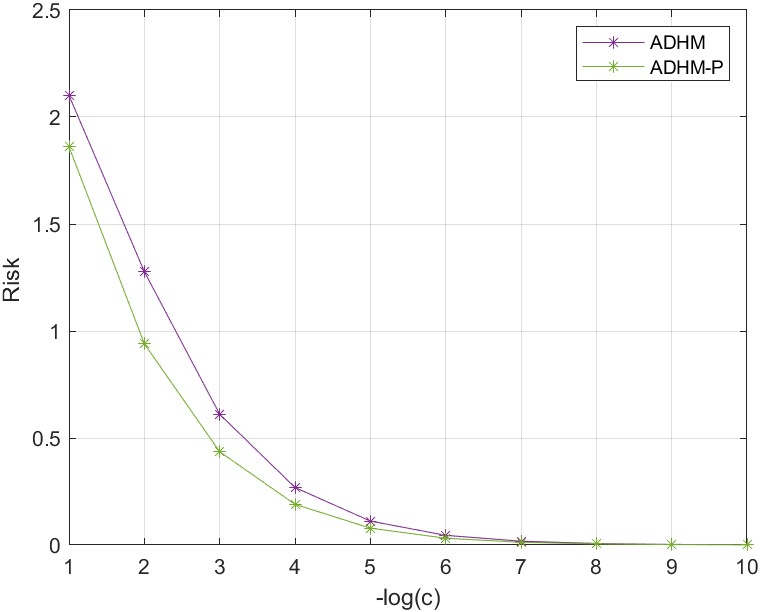}
    \caption{Performance comparison of two proposed algorithms: ADHM algorithm (ADHM), and ADHM with Predictive time scheduling (ADHM-P) for \( M = 5 \), \( K = 2 \), \( \lambda_f = 0.5 \), \( \lambda_g = 10 \), \( \alpha = \beta = 0.1 \), and threshold \( P_{\text{Th}} = 0.7 \). Results are based on \( 10^6 \) Monte Carlo trials. Shown: Bayes risk versus \( -\log c \).}
    \label{fig:RISK3}
\end{figure}

The performance results are illustrated in Fig.~\ref{fig:AD3}, Fig.~\ref{fig:AD_vs_PE3}, and Fig.~\ref{fig:RISK3}, which display the average detection delay, the error probability, and the Bayes risk, respectively. Each result is averaged over \( 10^6 \) independent Monte Carlo trials to ensure statistical robustness.

Across all three metrics, the ADHM-P algorithm demonstrates significant improvements over the original ADHM. By intelligently scheduling observations, ADHM-P reduces both the detection delay and error probability, while also achieving lower Bayes risk. These results clearly underscore the effectiveness of integrating predictive time scheduling into the anomaly detection framework.

The simulation results across a range of observation models, system sizes, and parameter settings collectively demonstrate the effectiveness and robustness of the proposed ADHM algorithm. In all scenarios, ADHM consistently outperforms existing benchmarks such as the deterministic DGF policy and the randomized Chernoff test, achieving lower detection delays and Bayes risks while maintaining tight control over the error probability. This performance advantage stems from ADHM’s principled use of the hidden Markov structure to guide adaptive sampling decisions.

Moreover, when predictive time scheduling is permitted, the extended ADHM-P algorithm exhibits further gains. By selectively sampling only when observations are expected to be informative—based on the evolving belief about the anomaly's hidden state—ADHM-P significantly reduces unnecessary sampling and improves overall efficiency. The results show that ADHM-P not only maintains the detection accuracy of ADHM but also achieves substantial improvements in delay and risk, highlighting the practical benefits of temporally adaptive strategies in resource-constrained or cost-sensitive settings.

\section{Conclusion}
\label{sec:Conclusion}

This work addressed the problem of detecting an anomalous process—referred to as the target—among \( M \) parallel processes (or cells), under a sampling constraint that allows observing only \( K \leq M \) cells at each time step. The main challenge lies in developing an active, sequential strategy for selecting which cells to observe in order to minimize the expected detection delay, while ensuring that the probability of detection error remains within acceptable bounds.

The observation model accounts for temporal dependencies through an HMM: observations from normal cells follow a fixed distribution \( f \), while those from the anomalous cell evolve according to an HMM that alternates between the same nominal distribution \( f \) and an abnormal distribution \( g \), depending on the hidden state. This structure introduces additional complexity into the decision-making process, as it requires accounting for both spatial uncertainty (which cell is anomalous) and temporal dynamics (evolution of hidden states).

To address this, we proposed the ADHM algorithm—a sequential strategy that adaptively selects the most promising cells based on a belief-driven scoring mechanism. The development of ADHM is grounded in the asymptotic optimality of the oracle case, where the true distribution parameters are assumed to be known. While ADHM operates under partial knowledge, it leverages structural insights from the oracle regime to guide practical decision-making. Extensive simulations demonstrate its consistent performance advantage over benchmark methods, including the deterministic DGF policy and the randomized Chernoff test.

We further extended the algorithm to incorporate predictive time scheduling, leading to the ADHM-P variant. This extension introduces a scheduling mechanism that determines not only which cells to observe, but also \emph{when} to observe, based on the predicted informativeness of future observations. Sampling is performed only when the estimated probability of being in the abnormal state exceeds a specified threshold, or when deferring observations becomes costlier than probing. This allows the algorithm to reduce unnecessary sampling without sacrificing detection accuracy.

Simulation results confirm that ADHM-P significantly reduces the average detection delay compared to the original ADHM, for the same level of error probability. These findings highlight the effectiveness of temporally adaptive sampling strategies in scenarios where observation costs or resource constraints are critical.

Overall, this study demonstrates the power of combining active hypothesis testing with hidden state modeling and adaptive scheduling, offering a principled and practical solution for efficient anomaly detection in dynamic and resource-limited environments.

\section{APPENDIX}
\label{sec:Appendix}

\subsection{Proof of Theorem 1}
As described earlier, in the oracle setting, it is assumed that for each sample $y(t)$, the algorithm is provided with the true value of $P^0_t$ by an oracle. Specifically, the distribution of sample $y$, given that the oracle provides $P^0_t = P_d$ is:
\begin{equation}
    \tilde{g}_d(y) = P_df(y)+  \left(1-P_d\right) g(y), \;\forall d=1, ..., D,
\end{equation}
where $\bar{P}=(P_1, ..., P_D)$ denotes a stationary distribution. This stationary distribution governs the behavior of the sampled observations from the abnormal process, with each $P_d$ corresponding to the abnormal distribution $\tilde{g}_d$.

We begin by introducing the notation used throughout the proof. Let $\textbf{1}_{\{P^0_t=P_d\}}$ denote an indicator function that equals $1$ when $P^0_t = P_d$, and $0$ otherwise, at time $t$. As a result, $\textbf{1}_{\{P^0_t=P_d\}}$ follows a Bernulli distribution with parameter $P_d$. 
With the above definitions, the LLR can be written as:
\begin{equation}
\ell_m(n) \triangleq \sum_{d=1}^{D}\textbf{1}_{\{P^0_n=P_d\}}\ell_{m,d}(n),
\end{equation}
where
\begin{equation}
\begin{aligned}
\label{LLR new}
\ell_{m,d}(n) \triangleq \log \frac{\tilde{g}_d(y_m(n))}{f(y_m(n))}.
\end{aligned}
\end{equation}
Now, the sum LLR is given by:
\begin{equation}
    S_m(n)=\sum_{t=1}^{n} \ell_{m}(t)\textbf{1}_m(t).
\end{equation}
The rate function is defined as:
\begin{equation}
\label{eq:rate_function}
\begin{aligned}
I^*(M,K)\triangleq \sum_{d=1}^{D} P_d I^*_d(M,K),
\end{aligned}
\end{equation}
where 
\begin{equation}
\begin{aligned}
I^*_d(M,K)\triangleq D(\tilde{g}_d||f)+\frac{K-1}{M-1}D(f||\tilde{g}_d)
\end{aligned}
\end{equation}
denotes the rate function corresponding to observations originating solely from the anomalous process governed by the distribution $\tilde{g}_d$. In the sequel, we will show that the asymptotically optimal Bayes risk achieved by ADHM is given by \( -c \log c / I^*(M, K) \) as $c\rightarrow 0$.

Let
\begin{equation}
\begin{aligned}
\label{def_N_j}
N_j(n) \triangleq \sum_{t=1}^n \textbf{1}_j(t)
\end{aligned}
\end{equation}
be the number of times cell $j$ has been observed up to time $n$.
We define
\begin{equation}
\begin{aligned}
\label{def_S_m,j}
\Delta S_{m,j}(n) \triangleq S_m(n) - S_j(n)
\end{aligned}
\end{equation}
as the difference between the observed sum LLRs of cell $m$ and $j$, and
\begin{equation}
\begin{aligned}
\label{def_Delta_S}
\Delta S_m(n) \triangleq \min_{j\neq m} \Delta S_{m,j}(n).
\end{aligned}
\end{equation}

For convenience, we define $\tilde{\ell}_{j}(i)$, a zero mean random variable under hypothesis $H_m$, as:
\begin{equation}
\begin{aligned}
\label{def_l_gal}
\tilde{\ell}_{j}(i) \triangleq
\begin{cases}
\ell_{j}(i) - \sum_{d=1}^{D}P_dD(\tilde{g}_d||f) & \text{if } j = m, \\
\ell_{j}(i) + \sum_{d=1}^{D} P_d D(f||\tilde{g}_d) & \text{if } j \neq m.
\end{cases}
\end{aligned}
\end{equation}

In this subsection, we show that the ADHM algorithm with oracle achieves the lower bound on the Bayes risk as $c \rightarrow 0$. We specifically focus on the case where $K \geq 2$ cells are observed at a time. Note that the case where $K = 1$ is simpler and can be proven with minor modifications.

\textbf{Step 1:} We begin by showing that the error probability of the ADHM algorithm is $O(c)$. Specifically, we next show that the error probability under ADHM is upper-bounded by:
\begin{equation}
\label{eq_Lemma_5}
P_e \leq (M-1)c.
\end{equation}

To prove this, let $\alpha_{m,j} = \textbf{P}_m(\delta=j)$ for all $j \neq m$. Therefore, $\alpha_m = \sum_{j \neq m} \alpha_{m,j}$. By the definition of the stopping rule in~\eqref{eq: stopping rule}, hypothesis \( H_j \) is accepted when \( \Delta S_j(\tau) \geq -\log c \), which in turn implies \( \Delta S_{j,m}(\tau) \geq -\log c \). Therefore, for all \( j \neq m \), we have:
\begin{equation}
\label{e_5_1}
\begin{aligned}
\alpha_{m,j} & = \textbf{P}_m(\delta = j) \\&
= \textbf{P}_m(\Delta S_j(\tau) \geq -\log c) \leq \textbf{P}_m(\Delta S_{j,m}(\tau) \geq -\log c) \\& \leq c \textbf{P}_j(\Delta S_{j,m}(\tau) \geq -\log c) \leq c. 
\end{aligned}
\end{equation}
where the change of measure in the second inequality follows from the fact that $\Delta S_{j,m} \geq -\log c$. Summing over all incorrect hypotheses, we obtain:
\begin{equation}
\label{e_5_2}
\alpha_{m} = \sum_{j \neq m}\alpha_{m,j} \leq (M-1)c.
\end{equation}
This completes Step 1. 

Next, recall that $S_m(n)$ is a random walk with a positive expected increment given by $\mathbf{E}_m(\ell_m(n)) = \sum_{d=1}^{D}P_d D(\tilde{g}_d||f) > 0$. In contrast, for any $j \neq m$, $S_j(n)$ is a random walk with a negative expected increment, $\mathbf{E}_m(\ell_j(n)) = -\sum_{d=1}^{D}P_dD(f||\tilde{g}_k) < 0$. Intuitively, this implies that, in the long run, the sum LLR under the true hypothesis will eventually dominate those of all other null hypotheses as $n$ grows large. We now proceed to formalize this behavior by quantifying the rate of separation. To that end, we define $\tau_1$ as the earliest time at which $S_m(n) \geq S_j(n)$ holds for all $j \neq m$, and such that this dominance holds for all $n \geq \tau_1$. 

\begin{definition}
$\tau_1$ is defined as the smallest integer after which $S_m(n) >  S_j(n)$ holds for all $j \neq m$ and for all $n \geq \tau_1$.
\end{definition}

\begin{remark}
By saying that the ADHM algorithm is implemented indefinitely, we refer to the case where it continues probing the cells according to its selection rule, without applying the stopping rule.
\end{remark}

\textbf{Step 2:} Next, we show that $\tau_1$ is small with high probability. Specifically, assume that the selection rule of the ADHM algorithm is implemented indefinitely. Then, there exist $C > 0$ and $\gamma > 0$ such that
\begin{equation}
\label{e_Lemma_7}
\textbf{P}_m(\tau_1 > n) \leq Ce^{-\gamma n},
\end{equation}
for $m=1,\ldots,M$.

To prove this, note that:
\beq\label{eq:tau1}
\bea{l}
\displaystyle\mathbf{P}_m\left(\tau_1>n\right)\leq\mathbf{P}_m\left(\max_{j\neq m}\;\sup_{t\geq n}\;\left(S_j(t)-S_m(t)\right)\geq 0 \right) \vspace{0.2cm} \\ \hspace{2cm}
\leq\displaystyle\sum_{j\neq m}\;\sum_{t=n}^{\infty}\mathbf{P}_m\left(S_j(t)\geq S_m(t)\right)\;.
\ena
\eeq
Therefore, it suffices to show that there exist $C>0$ and $\gamma>0$ such that $\mathbf{P}_m\left(S_j(n)\geq S_m(n)\right)\leq Ce^{-\gamma n}$.

Next, we bound each term in the summation on the RHS of (\ref{eq:tau1}). Let $\rho>0$ denote a fixed small constant. Thus,
\beq
\bea{l}
\label{eq:l_tau_1_policy1_Sj_geq_Sm}
\mathbf{P}_m\left(S_j(n)\geq S_m(n)\right) \vspace{0.2cm} \\ %\hspace{0.3cm}
\leq\mathbf{P}_m\left(S_j(n)\geq S_m(n), N_j(n)<\rho n , N_m(n)<\rho n\right)
\vspace{0.2cm} \\ %\hspace{0.3cm}
+\mathbf{P}_m\left(S_j(n)\geq S_m(n), N_j(n)\geq \rho n\right) \vspace{0.2cm} \\ %\hspace{0.3cm}
+\mathbf{P}_m\left(S_j(n)\geq S_m(n), N_m(n)\geq \rho n\right).
\ena
\eeq
Since $N_j(n) \geq \rho n$ and $N_m(n) \geq \rho n$ hold for the second and third terms on the RHS, respectively, we can directly apply the Chernoff bound to each. This yields the existence of constants $\gamma_1 > 0$ and $C_1 > 0$ such that both terms are upper bounded by $C_1 e^{-\gamma_1 n}$. It thus remains to show that the first term on the RHS also decays exponentially with $n$. Observe that the event $(N_j(n) < \rho n,\, N_m(n) < \rho n)$ implies that at least $\tilde{n} = n - N_j(n) - N_m(n) \geq n(1 - 2\rho)$ time steps occur in which neither cell $j$ nor cell $m$ is probed. This scenario also requires that a random walk with a negative expected increment, namely $S_r(n)$, dominates both $S_j(n)$ and $S_m(n)$, even though $S_j(n)$ and $S_m(n)$ are probed only a small fraction of the time, while $S_r(n)$ is probed frequently. By applying a similar Chernoff bound argument, we conclude that there exist constants $\gamma_2 > 0$ and $C_2 > 0$ such that
$\mathbf{P}_m\left(S_j(n) \geq S_m(n)\right) \leq C_2 e^{-\gamma_2 n}$ which completes Step 2.

\textbf{Step 3:} We now define a second random time, $\tau_2 \geq \tau_1$, representing the point at which sufficient information has been accumulated to distinguish the true hypothesis from at least one false hypothesis. In this step, we show that in the asymptotic regime, the time interval between $\tau_1$ and $\tau_2$, denoted by $n_2$, satisfies $
n_2 \sim \frac{-\log c}{I^*(M, K)}$.

\begin{definition}
$\tau_2$ denotes the smallest integer such that 
$\sum_{i=\tau_1}^n\ell_m(i)\boldsymbol{1}_m(i)\geq -\frac{ \sum_{d=1}^D P_d D(\tilde{g}_d||f)}{I^*(M,K)}\log c$ and
$\sum_{i=\tau_1}^n\ell_{j_n}(i)\boldsymbol{1}_{j_n}(i)\leq \frac{(K-1) \sum_{d=1}^{D} P_d D(f||\tilde{g}_d)}{(M-1)I^*(M,K)}\log c$ for some $j_n \neq m$, for all $n \geq \tau_2 \geq \tau_1$.
\end{definition}

\begin{definition}
 Define $n_2 \triangleq \tau_2 - \tau_1$ as the total time between $\tau_1$ and $\tau_2$.
\end{definition}

Specifically, in this step we prove the following. Assume that the ADHM algorithm is implemented indefinitely. Then, for every fixed $\epsilon >0$, there exist $C >0$ and $\gamma >0$ such that
\begin{equation}
\label{Lemma_8_eq}
\textbf{P}_m(n_2 > n) \leq Ce^{-\gamma n} \quad \forall n >-(1+\epsilon)\log c/ I^*(M,K),
\end{equation}
for all $m=1,\ldots,M$.

To prove this, let $D'(f||\tilde{g}_k) = (K-1)D(f||\tilde{g}_k)/ (M-1)$. Then, $I^*(M,K) = \sum_{d=1}^{D} P_d D(\tilde{g}_k||f)+  \sum_{d=1}^D P_d D'(f||\tilde{g}_k)$ and cell $m$ is observed for all $n$, since $n \geq \tau_1$ and $S_m(n) > S_j(n)$ for all $j \neq m$. Let
\begin{equation}
\label{e8_6}
\begin{aligned}
N'_j(\tau_1+t) \triangleq \sum_{i=\tau_1+1}^{\tau_1+t} \textbf{1}_j(i)
\end{aligned}
\end{equation}
for $j \neq m$. 

Let $j^*(\tau_1+t)= \arg\max_{j \neq m}N'_j(\tau_1+t)$ be the cell index that was observed the largest number of times since $\tau_1$ has occurred up to time $\tau_1+t$.
Note that if $\sum_{i=\tau_1}^n \ell_{j^*(\tau_1+t)}(i)\textbf{1}_{j^*(\tau_1+t)}(i)\leq \frac{\sum_{d=1}^{D} P_d D'(f||\tilde{g}_d)}{I^*(M,K)}\log c$ and $\sum_{i=\tau_1}^n\ell_{m}(i)\textbf{1}_m(i)\geq -\frac{\sum_{d=1}^{D} P_d D(\tilde{g}_d||f)}{I^*(M,K)}\log c$ for all $t \geq n$, then $n_2 \leq n$. Hence,
\begin{equation}
\label{e8_7}
\begin{aligned}
&\textbf{P}_m(n_2 > n) \\&
 \leq  \textbf{P}_m\left(\inf _{t \geq n}  \sum_{i=\tau_1+1}^{\tau_1+t}\ell_{m}(i)\textbf{1}_m(i)\leq -\frac{\sum_{d=1}^{D}P_d D(\tilde{g}_d||f)}{I^*(M,K)}\log c\right) \\&
 + \textbf{P}_m \left( \sup_{t\geq n} \sum_{i=\tau_1+1}^{\tau_1+t} \ell_{j^*(\tau_1+t)}(i)\textbf{1}_{j^*(\tau_1+t)}(i)\right.
 \\&\hspace{4.5cm}
 \left.
 \geq \frac{\sum_{d=1}^{D} P_d D'(f||\tilde{g}_d)}{I^*(M,K)}\log c\right).
\end{aligned}
\end{equation}
For the second term in the RHS of \eqref{e8_7}, note that $N'_{j^*(\tau_1+t)}(\tau_1+t) \geq \frac{(K-1)t}{M-1}$, since $t$ observations are taken from cell $m$ and $(K-1)t$ observations are taken from $M-1$ cells (for $j \neq m$). Thus, for any $\epsilon >0$ there exists $\epsilon_2 = \frac{\sum_{d=1}^{D}P_d D(f||\tilde{g}_d)\epsilon}{1+\epsilon} >0$ such that:
\begin{align}
&\sum_{i=\tau_1+1}^{\tau_1+t} \ell_{j^*(\tau_1+t)}(i)\, \mathbf{1}_{j^*(\tau_1+t)}(i) 
- \frac{\sum_{d=1}^{D} P_d D'(f \,\|\, \tilde{g}_d)}{I^*(M, K)} \log c \notag \\
&\quad \leq \sum_{i=\tau_1+1}^{\tau_1+t} \tilde{\ell}_{j^*(\tau_1+t)}(i)\, \mathbf{1}_{j^*(\tau_1+t)}(i) \notag \\
&\qquad - N'_{j^*(\tau_1+t)}(\tau_1+t) 
\sum_{d=1}^{D} P_d D(f \,\|\, \tilde{g}_d) \notag \\
&\qquad - \frac{\sum_{d=1}^{D} P_d D'(f \,\|\, \tilde{g}_d)}{I^*(M, K)} \log c \notag \\
&\quad \leq \sum_{i=\tau_1+1}^{\tau_1+t} \tilde{\ell}_{j^*(\tau_1+t),k}(i)\, 
\mathbf{1}_{j^*(\tau_1+t)}(i) \notag \\
&\qquad - \frac{(K-1)t}{M-1} 
\sum_{d=1}^{D} P_d D(f \,\|\, \tilde{g}_d) \notag \\
&\qquad - \frac{(K-1)}{(M-1) I^*(M, K)} 
\sum_{d=1}^{D} P_d D(f \,\|\, \tilde{g}_d) \log c \notag \\
&\quad = \sum_{i=\tau_1+1}^{\tau_1+t} \tilde{\ell}_{j^*(\tau_1+t),k}(i)\, 
\mathbf{1}_{j^*(\tau_1+t)}(i) \notag \\
&\qquad - \frac{(K-1)t}{M-1} \sum_{d=1}^{D} P_d D(f \,\|\, \tilde{g}_d) 
\left[ 1 + \frac{-\log c}{t \cdot I^*(M, K)} \right] \notag \\
&\quad \leq \sum_{i=\tau_1+1}^{\tau_1+t} \tilde{\ell}_{j^*(\tau_1+t),k}(i)\, 
\mathbf{1}_{j^*(\tau_1+t)}(i) - t \epsilon_2.
\label{eq:e8_8}
\end{align}
for all $t\geq n >-(1+\epsilon)\log c / I^*(M,K)$. Next, applying the Chernoff bound yields the desired exponential decay. A similar argument, with minor modifications, applies to the first term on the RHS of~\eqref{e8_7}. This completes Step 3.

\textbf{Step 4:}
What remained to show for the upper bound on the expected detection time, is that  the time interval between $\tau_2$ and the stopping time $\tau$ decays exponentially with $n$. This is a consequence of the design of the ADHM algorithm, which always selects cells with the highest sum LLRs, thereby ensuring that the dynamic range between the sum LLRs remains sufficiently small. Establishing this formally involves applying similar Chernoff bound techniques as those used in earlier parts of the analysis. As a result, the expected detection time $\tau$ under the ADHM algorithm is upper-bounded by: $\textbf{E}_m(\tau) \leq -(1+o(1))\frac{\log c}{I^*(M,K)}$ for all $m=1,2,\ldots,M$. Finally, combining the four steps with the asymptotic lower bound on the Bayes risk \cite{nitinawarat2013controlled}, completes the proof.

\subsection{Proof of Theorem 2}

We first establish the result for samples originating from the anomalous process. Let $\{P_d\}_{d=1}^D$ be a probability distribution, and let $\{g_d(y)\}_{d=1}^D$ be probability density functions. Define the mixture:
\[
g(y) = \sum_{d=1}^D P_d g_d(y).
\]
Let $f(y)$ be the distribution of the normal processes. To show that $I^*_{\mathrm{ADHM}} \geq I^*_{\mathrm{Chernoff}}$, we thus need to prove that:
\[
\sum_{d=1}^D P_d D(g_d \| f) \ge 
D\left( \sum_{d=1}^D P_d g_d \middle\| f \right).
\]
Note that the LHS can be written as:
\begin{equation}
\sum_{d=1}^D P_d \int g_d(y) \log \left( \frac{g_d(y)}{f(y)} \right) dy, \label{eq:left}
\end{equation}
and the RHS can be written as:
\begin{equation}
\int g(y) \log \left( \frac{g(y)}{f(y)} \right) dy. \label{eq:right}
\end{equation}

Subtracting \eqref{eq:right} from \eqref{eq:left} yields:
\begin{align}
\Delta 
&= \sum_{d=1}^D P_d \int g_d(y) \log\left( \frac{g_d(y)}{f(y)} \right) dy 
\\&\hspace{0.5cm}
- \int g(y) \log\left( \frac{g(y)}{f(y)} \right) dy \notag \\
&= \sum_{d=1}^D P_d \int g_d(y) 
\left[ 
\log\left( \frac{g_d(y)}{g(y)} \right) 
+ \log\left( \frac{g(y)}{f(y)} \right) 
\right] dy \notag \\
&\quad - \int g(y) \log\left( \frac{g(y)}{f(y)} \right) dy \notag \\
&= \sum_{d=1}^D P_d \int g_d(y) \log\left( \frac{g_d(y)}{g(y)} \right) dy \notag \\
&\quad + \sum_{d=1}^D P_d \int g_d(y) \log\left( \frac{g(y)}{f(y)} \right) dy \notag \\
&\quad - \int g(y) \log\left( \frac{g(y)}{f(y)} \right) dy \notag \\
&= \sum_{d=1}^D P_d \int g_d(y) \log\left( \frac{g_d(y)}{g(y)} \right) dy \notag \\
&= \int \sum_{d=1}^D P_d g_d(y) \log\left( \frac{g_d(y)}{g(y)} \right) dy \notag \\
&= \int g(y) \sum_{d=1}^D \frac{P_d g_d(y)}{g(y)} 
\log\left( \frac{g_d(y)}{g(y)} \right) dy \notag \\
&= \int g(y) D(P(d \mid y) \| P_d) dy \ge 0, \label{eq:delta}
\end{align}
where $P(d \mid y) = \frac{P_d g_d(y)}{g(y)}$ is a posterior over $d$ given $y$.

Since KL divergence is always non-negative, we conclude that:
\[
\sum_{d=1}^D P_d D(g_d \| f) \ge D(g \| f),
\]
with equality if and only if $g_d(y) = g(y)$ for all $d$ where $P_d > 0$. A similar line of reasoning applies to the normal processes, leading to the conclusion that the rate function under ADHM is always greater than that under the Chernoff test.
\hfill$\blacksquare$

\bibliographystyle{ieeetr}
%\bibliography{Bibliography}

\end{document}